 \documentstyle[aps,pra,multicol]{revtex}
\begin{document}
\def\ibd {{\it ibid. }}
\def\bfj{{\bf j}}
\def\bfq{{\bf q}}
\def\bfk{{\bf k}}
\def\bfrho{{\bf \rho}}
\def\om{\omega}
\def\tilg{\tilde{g}}
\def\tilg_{\tilde{g_-}}
\def\vare{\varepsilon}
\preprint{CNAT-99}
\title{Bose-Einstein condensation in a two-dimensional trap}
\author{Sang-Hoon Kim,$^1$  Changyeon Won,$^2$ Sung Dahm Oh,$^3$,
and Wonho Jhe,$^2$}
\address{$^1$Division of Liberal Arts, Mokpo National Maritime University,
 Mokpo 530-729, Korea}
\address{$^2$CNAT and Department of Physics, Seoul National University,
 Seoul 151-742, Korea}
\address{$^3$Department of Physics, Sookmyung Women's University, Seoul
 140-742, Korea}
\date{\today}
\maketitle
\begin{abstract}
The theory of Bose-Einstein condensation in a two-dimensional(2D)
harmonic trap is developed from  2D Gross-Pitaevskii equation. 
The 2D interaction strength  is obtained from a 2D  collision theory.
 We show the realization of 2D condensation 
 of trapped Bose atoms directly  by obtaining the stable solutions
for the  condensate wave function
 from the 2D Gross-Pitaevskii equation.
We calculate the ground-state energy of the 2D system,
 and also the wave function of  the 2D vortex state. 
In particular, the 2D energy state becomes less stable than
the 3D case with the number of trapped atoms.
The results of the 2D Bose condensation are also compared with 
those of the well-known 3D case.
\end{abstract}
\draft
 \begin{multicols}{2}
\section{introduction}
 The recent  experimental realization of Bose-Einstein Condensation(BEC)
 of alkali atoms in magnetic traps \cite{ande,brad,davi} has generated much
interest and activity in theoretical and experimental physics.
An interesting, but less addressed, question is whether such a phase 
transition due to global coherence exists in two-dimensional(2D) space.
	Even though it has been known that 2D 
BEC of a uniform(untrapped) Bose gas cannot 
occur in 2D at a finite temperature since thermal fluctuations 
destabilize the condensate \cite{hohe,destab},
    it has been suggested that spatially-varying potentials which
break the uniform distribution may create BEC in 2D inhomogeneous systems
 \cite{bagn,temp,jack,bayi,haug,gonz}.
In the presence of trapping potential, the effect of thermal fluctuations
are strongly quenched due to the different behavior exhibited
by the density of states.

At zero temperature, the 2D condensation can be described 
by the 2D Gross-Pitaevskii equation(GPE), 
a mean-field approximation for the macroscopic wave function
of weakly interacting bosons.   
	The GPE  is obtained from mean-field many-body quantum-statistical 
theory and has been proven to  explain the condensate
 state of dilute Bose system satisfactorily.
Although no direct experimental observation of 2D BEC has 
been reported yet,
 recent theoretical works suggest a possibility of 2D BEC
  in a trapped condition.  However, most of
  the theoretical approaches for 2D BEC have been rather indirect
and sometimes even  faced with several difficulties.

Tempere and Devreese \cite{temp} studied harmonically-interacting
bosons in 2D.  They calculated the critical temperature from a 
grand-partition function to show the occurrence of 2D BEC,
but without any knowledge of the 2D condensate wave function
 for the density  profile.
	Jackson {\it et al.} \cite{jack} studied the 2D vortex state
 by direct substitution of the 3D interaction strength
 $g_{3D} =  4\pi \hbar^2 a /m$
($a$ is the s-wave scattering length in 3D and $m$ is the atomic mass)
into the 2D GPE, which resulted in a dimensional inconsistency.
	 Bayindir {\it et al.} \cite{bayi} solved the 2D GPE using
 the two-fluid model  a density estimation and obtained the temperature dependence of the
 internal energy and condensate fraction, where
the 2D interaction strength was still treated just as a free variable.
	 Haugset {\it et  al.} \cite{haug} calculated the density 
profile and ground state energy for a finite number of 2D Bose particles 
by diagonalizing the Hamiltonian numerically, where a modified 
2D interaction strength was used to solve the dimensional inconsistency.
 	Analytical approach of interacting Bosons in 2D trap
by  2D nonlinear GPE was attempted by Gonzalez {\it et. al.} \cite{gonz},
and they  studied the ground-state energy density by applying
the Haugset {\it et al.}'s  modified form of the 2D interaction strength
to the 2D GPE. 

Previous 2D researches are based on the size-independent concepts:
if the atomic motion in any one axis is frozen completely in the ground state, 
the system is considered as two-dimension.  
Moreover, most of the studies considered the 2D trap as an extremely 
anisotropic 3D trap  and still far from the pure 2D system.
On the other hand, we will focus on the 2D system which is
confined by a 2D harmonic trap in $(x,y)$-direction and by ideal rigid 
walls in $z$-direction \cite{kim2}.  In this system, the 2D interaction
 strength has a logarithmic form which will be derived 
using a 2D scattering theory in the next section.
  Quasi-2D atomic systems, where atoms follow the 2D kinematics with the 3D
interactions \cite{kagan}, have been recently realized for spin-polarized 
atomic hydrogen adsorbed on the surface of liquid helium 
\cite{mosk,safo} and quasi-condensation has been indirectly demonstrated
\cite{safo}.
 Although no pure 2D BEC  has been studied experimentally yet, the 2D 
GPE is quite valid even in quasi-2D at low temperatures \cite{shly}.
Moreover it will provide us a guide to a better understanding of the 2D
 statistical properties of the quasi-2D system.

In this paper, we will solve the 2D GPE numerically 
and show the existence of the 2D BEC 
directly by obtaining the stable condensate wave functions  and
 the 2D ground state energy density of the trapped Bose atoms.
Possible vortex states in 2D are also discussed.   
The results of the 2D Bose condensation are then compared with 
those of the well-known 3D cases.

\section{two-dimensional scattering theory}
 The relation between interaction strength, 
$g$, and the s-wave scattering length, $a$, is  well-known in 3D  as
$g =  4\pi \hbar^2 a /m$.
However,  the 3D result
is  not applicable to a 2D GPE:  substitution of the 3D relation into 
the 2D GPE results in dimensional inconsistency.
As a first step to obtain the correct 2D GPE,
the interaction strength in 2D is derived from the 
following 2D scattering theory \cite{schi,kim,verh,kaga}.
We begin with the 2D time-independent Schr\"{o}dinger equation 
with the interaction $U(\rho)$, given by
\begin{equation}
(\nabla^2_\rho + k^2 ) \psi_k(\bfrho) = U(\rho) \psi_k(\bfrho),
\label{1}
\end{equation}
where $k^2=2\mu_m E/\hbar^2$.
$\mu_m$ is the reduced mass of two particles ($m/2$).
One can find the general solution of Eq. (\ref{1}) with the help
of the 2D Green's function
\begin{equation}
(\nabla^2_\rho + k^2 ) G_k(\bfrho,\bfrho') = \delta^2({\bf\rho} - \bfrho').
\label{3}
\end{equation}

The general solution of the 2D Green's function is the Hankel
 function \cite{mors,arfk}.
Therefore the  2D  wave function can be expressed as
\begin{eqnarray}
\psi_k(\bfrho) &=& e^{i\bfk\cdot\bfrho} + \int d^2\rho'\,
 G_k(\bfrho,\bfrho') U(\rho')\psi(\rho')
\nonumber \\
&=&  e^{i\bfk\cdot\bfrho} - \frac{i}{4}\int d^2\rho' \,
 H_0(k|\bfrho - \bfrho'|) U(\rho')\psi(\rho')
\nonumber \\ 
&\simeq& e^{i\bfk\cdot\bfrho} 
- \frac{i}{2} \frac{e^{i(k \rho -\frac{\pi}{4})}}{\sqrt{2\pi k \rho}}
\int d^2\rho' \, U(\rho') e^{i(\bfk-\bfk')\cdot\bfrho'}.
\label{7}
\end{eqnarray}
Here $H_0$ is the first kind Hankel function of order zero  defined by
$H_0(x)=J_0(x) + i N_0(x)$, where
  $J_0$  and $N_0$ are  Bessel  and  Neumann function of order zero.
For large $\rho$, it has the asymptotic behavior
$H_l(x) \simeq \sqrt{2/\pi x}\, e^{i(x-l\pi/2 -\pi/4)}$,
 where $l$ is an integer.  The Born approximation is applied to
 the last step, and $\bfk' = k \hat{\bfrho}$.

In 2D, for large $\rho$,
 the scattered wave function should have the form
\begin{equation}
 \psi_k(\bfrho) \simeq  e^{i\bfk\cdot\bfrho} +
F_k \frac {e^{i(k \rho -\frac{\pi}{4})}}{\sqrt{\rho}},
\label{9}
\end{equation}
where  $F_k$ is the  first-order Born scattering amplitude 
in 2D which has a dimension of $(length)^{1/2}$.
This asymptotic formula  should be a solution of the time
independent Schr\"{o}dinger equation (\ref{1}), and satisfy 
$\rho \rightarrow \infty$ limit of the Green's function solution.
The scattering cross-section in 2D now has  the dimension of
 {\it length} given by
\begin{equation}
\frac{d \sigma_{2D}}{d\theta} = 
\frac{|\bfj_{sc}| \rho}{|\bfj_{inc}|} = |F_k|^2,
\label{11}
\end{equation}
where $\bfj_{inc}$ and $\bfj_{sc}$ are incident and scattered flux
 densities.
We note that the total scattering cross-section for small $k$
 should be $2\pi |F_k|^2$ in 2D instead of $4\pi |f_{3D}|^2$ in 3D. 
	If one assumes  the delta-function
 type interaction  $U(\rho)=(m g/\hbar^2) \delta^2(\rho),$
 the 2D scattering amplitude $F_k$  is obtained in the complex form 
 from Eqs. (\ref{7}) and (\ref{9}) as 
\begin{equation}
F_k  =  -\frac{i}{2}\frac{1}{\sqrt{2\pi k}}\frac{m g}{\hbar^2}. 
\label{131}
\end{equation}

The next step is to find the relation between the scattering 
amplitude  and  the scattering length in 2D.
The  well-known relation  $f_{3D} = -a$, where $f_{3D}$ is the 3D 
scattering amplitude and $a$ is the s-wave scattering length in 3D, 
is not directly applicable in 2D.
In general, the s-wave scattering length in 2D, $b$,
  is different from the one in 3D, $a$, and 
the s-wave scattering length in 2D is not known yet from experiment.

The relation is obtained from partial wave analysis  of collision theory.
The incident wave can be written as an expansion of plane wave 
in cylindrical coordinate \cite{arfk}
\begin{equation}
e^{i k \rho \cos\theta} = \frac{1}{2}\sum_{l=-\infty}^{\infty}
i^l \left[ H_l(k \rho) + H_l^\ast(k \rho)\right] e^{i l \theta},
\label{14}
\end{equation}
where $l$ is an integer.
	Away from the range of potential,
the scattered wave has a phase shift $\delta_l$, and the scattering matrix
is obtained as $S_l = e^{2i\delta_l}$ since $|S_l|=1$ for elastic scattering.
Therefore, for large $\rho$,  the solution of Eq. (\ref{1})
is written as
\begin{eqnarray}
\psi_k(\bfrho) &=& e^{i\bfk\cdot\bfrho}+ \frac{1}{2}\sum_{l=-\infty}^{\infty}
i^l  (S_l-1) H_l(k \rho)  e^{i l \theta}
\nonumber \\
&\simeq & e^{i\bfk\cdot\bfrho}+  \frac{1}{\sqrt{2 \pi k \rho}}
\sum_{l=-\infty}^{\infty} i^l (e^{2i\delta_l}-1)
e^{i(k\rho - \frac{l\pi}{2} -\frac{\pi}{4})} e^{i l \theta}.
\label{16}
\end{eqnarray}

Scattering length is defined as the distance where the two-body wave 
function vanishes for zero energy.  The phase shift due to potential scattering
 is given as a function of the scattering length, and  
it is well-known as $\delta_{0,3D} = - k a$ in 3D, whereas
\begin{equation}
 \delta_0  = \frac{\pi}{2} \frac{1}{\ln k b}\left[ 1 
+  {\cal O} \left( \frac{1}{\ln kb}\right) \right]  \; ,
\label{163}
\end{equation}
in 2D \cite{verh}. Note that it is effective only in the
low energy scattering limit, $k b << 1$.
It can be easily checked as follows.
The scattering length is just the intercept of the radial wave function
satisfying the boundary condition $\psi_k(b) = 0$.
  Therefore, for $l=0$, the wave function at large distance
and small $k$ is expressed as 
\begin{eqnarray}
\psi_k(\bfrho)&=&J_0(k \rho) - \frac{J_0(k b)}{N_0(k b)} N_0(k \rho)
\nonumber \\
&\simeq& \frac{2}{\sqrt{2 \pi k \rho}}
\left[\cos \left(k\rho - \frac{\pi}{4}\right)
- \frac{\pi}{2 \ln kb} \sin \left(k\rho - \frac{\pi}{4} \right)\right]
\nonumber \\
& = &  \frac{2}{\sqrt{2\pi k \rho}}
\cos\left(k\rho -\frac{\pi}{4} +\delta_0\right).
\label{161}
\end{eqnarray}
The necessary condition for the validity of the Born approximation
is that the phase shift $\delta_0$ be very small for small $k$, which can be
easily confirmed  from Eq. (\ref{163}).
Note that unlike the 3D case where $a$ can be negative,
we do not consider the negative scattering length here
since the centrifugal potential of the lowest partial wave is negative in 2D
so that the extrapolated local wave function cuts the radial axis
always above the origin \cite{verh}.

Comparing Eq. (\ref{16}) with Eq. (\ref{9}), 
 the scattering amplitude is written as a series expansion, and
 the $l=0$ state contributes to the 2D system.  Therefore, the 2D scattering
amplitude becomes 
\begin{eqnarray}
F_k &=& \frac{1}{\sqrt{2\pi k}}
 \sum_{l=-\infty}^{\infty} \left( e^{2i\delta_l}-1 \right) e^{i l\theta}
\nonumber \\
 & = & \frac{2i \delta_0}{\sqrt{2\pi k}} (1 + i \delta_0 + ...)
\nonumber \\
 & \simeq & \frac{i \pi}{\sqrt{2\pi k}}\frac{1}{\ln k b} .
\label{17}
\end{eqnarray}

Finally, we obtain the 2D  interaction $g$
from Eqs. (\ref{131}) and (\ref{17}) as
\begin{equation}
g =  - \frac{2 \pi \hbar^2}{m}\frac{1}{\ln k b}.
\label{201}
\end{equation}
In general the 2D interaction strength  is given as \cite{schi}
\begin{equation}
g = \frac{ 4\pi \hbar^2 \xi}{m} \; .
\label{5}
\end{equation}
Here $\xi$ is a dimensionless atomic parameter given by
 \begin{equation}\xi  =  -\frac{1}{2}  \frac{1}{\ln k b} \; ,
\label{6}\end{equation}
where $\xi$ is positive since  $k b \ll 1$.
Note that the logarithmic dependence of the interaction term
suggests  similar condensate characteristics in 2D
for most bosonic alkali atoms with positive scattering length.

In addition to  the 2D scattering length, $b$, 
we do not have any reliable value of the wave-vector, $k$, in 2D.
An approximation that the wave-vector $k$  be  the inverse of the largest distance 
available in the perpendicular direction may not be correct. 
However, the experimental results provide the value of the 
product $kb$ as described in the next section.
Note that in an extremely anisotropic 3D system, 
the dimensionless atomic parameter is given as
$\xi_{e} = (1/\sqrt{2 \pi}) a/a_z,$
 where $a_z = \sqrt{\hbar/m \omega_z}$,  
and moreover $\xi_{e} < \xi$ in general.

\section{the condensate state and its energy state}
Now the 2D condensate wave function of trapped dilute Bose atoms of mass
 $m$  can be obtained from the 2D GPE
\begin{equation}
\left[ -\frac{\hbar^2}{2m}\nabla^2_\rho + V_{ext}(\bfrho)
+ N g \psi^2(\bfrho) \right] \psi(\bfrho)= \mu \psi(\bfrho) \; ,
\label{21}
\end{equation}
where $\int d^2\rho \, \psi^2(\rho) = 1$,
 and $N$ is the number of condensate particles.
Here, we assume a 2D isotropic, harmonic trap
$V_{ext}(\rho) = \frac{1}{2} m \omega^2 \rho^2$
 where $\omega$ is the trap frequency.
Then we can simplify Eq. (\ref{21}) for numerical calculation
by introducing  dimensionless variables
($\bfrho \rightarrow a_{ho} \bfrho,$ $ \mu \rightarrow \hbar\omega\mu$,
 and $\psi \rightarrow a_{ho}^{-1}\psi$)
\begin{equation}
\left[  -\nabla^2_\bfrho +\bfrho^2 -2\mu +
8\pi \xi N  \psi^2(\bfrho)\right] \psi(\bfrho)=0,
\label{8}
\end{equation}
where $a_{ho} = \sqrt{\hbar/m\omega }$ is the 2D harmonic oscillator length,
 and $\mu$ is the 2D chemical potential which is obtained from 
the normalization condition. 
Note that the product of incident wave-vector and scattering length,
$kb$, is the only atomic parameter that contributes to the condensate states.

In the numerical calculation of the GPE, we need  the value of the 
atomic parameter  $\xi$ in Eq. (\ref{5}) which is a function of
$k$ and  $b$. Although they are not known separately, we are able to deduce their product
$kb$ from experimental data, which appears together in the 2D GPE.
It is certain that different Bose atoms have different 2D scattering lengths,
but the logarithmic dependence makes the difference less sensitive.
A recent experiment of the hydrogen on helium surface by Safonov {\it et al.} \cite{safo} 
has reported $\xi = 1/7$ which indicates $k b = 3 \times 10^{-2}$.
Although their system actually satisfies a quasi-2D condition, one can still 
take this value in the 2D GPE as an effective interaction potential. 
Since it is well known that the 2D 
GPE is valid even in quasi-2D at low temperatures \cite{shly}, and 
moreover it hints the 2D atomic characteristics in the 
quasi-2D system, it is useful to quantify the criterion for the quasi-2D, or the
effective thickness of the 3D system to exhibit the 2D statistical properties.
The criteria of the effective thickness of the  trapped 2D Bose system 
may by obtained \cite{kim2}.

	The next procedure to solve  Eq. (\ref{8}) is almost
similar to those of 3D \cite{dalf,edwa}.
We have plotted the 2D condensate wave functions versus $\rho$
for several values of atom number $N$ in Fig. 1. 
It corresponds to the $z=0$ cut of the
anisotropic contour plot of  the ground states in 3D. 
The spatial distribution of the condensate in 2D is much broader
 than that in 3D,
and the condensate wave functions approach the parabolic limit more rapidly
with the increase of the number of atoms.
In other words, the effect of atomic interaction potential
becomes more prominent in 2D.

In the non-interacting case, the solution is still Gaussian with
$\psi(\rho) = \pi^{-1/2} e^{-\rho^2/2}$.
In the strongly repulsive limit or Thomas-Fermi limit,
 it has the parabolic solution of 
$\psi^2(\rho) = (2\mu - \rho^2)/8 \pi \xi N$.
The overall shape of the 
condensate wave functions are similar with that of 3D, but it approaches
the parabolic limit more quickly with the number of atoms \cite{dalf}. 
  Therefore, the peak of the 
density profile decreases much faster in 2D.
The 2D healing length that balances between the quantum pressure
and the interaction energy of the condensate is also different from that
of 3D.   
Refer to TABLE 1 for detailed comparison. 

The ground-state energy 
for 2D condensate bosons  can be calculated in a similar way.
 With the dimensionless variables defined before,
we obtain the dimensionless energy density as 
\begin{eqnarray}
\vare  &=& \int \left[ \frac{1}{2}|\nabla \psi|^2
+ \frac{1}{2}\rho^2 |\psi|^2 + 2 \pi \xi N |\psi|^4 \right]d^2\rho.
\label{12}
\end{eqnarray}
        Using a Gaussian trial function, we easily find 
the ground-state energy density satisfies
\begin{eqnarray}
\vare \ge \sqrt{1 + 2 \xi N}.
\label{13}
\end{eqnarray}
The ground-state energy per particle in Eq. (\ref{13}) is plotted in Fig. 2, 
and  compared with the well-known 3D results of  $^{87}$Rb.
The 2D system becomes more excited for a given $N$ and less  
stable as $N$ is increased with respect to the 3D case. 
We have summarized the fundamental differences between our 2D results
and the well-known 3D ones in TABLE 1.

\section{the vortex state}
Now let us consider the vortex states of the 2D system.
The hydrodynamic theory connected to superfluidity is understood
by vortex. The 2D system can be rotating about the center of the 2D trap
to give quantized circulation of atomic motion. 
The angular momentum quantum number $\kappa$ gives the quantum winding
of the 2D vortex state.  
A vortex state with winding number $\kappa$ is written as
$\psi(\bfrho) = \phi(\bfrho) e^{iS(\bfrho)}$,
where $\phi(\bfrho) = \sqrt{n(\bfrho)}$ is the modulus. The phase $S$
is chosen as $\kappa \theta$ where $\kappa$ is an integer.
An angular momentum quantum number $\kappa$ can be then assigned to the quantum 
winding of the 2D vortex state, and
one finds the vortex states with a tangential velocity 
$v = \kappa \hbar/m \rho$. 
As a result of the  quantum circulation, the angular momentum of the system 
with respect to the $\rho = 0$ axis 
becomes  $L = N \kappa \hbar$. 

        Adding the vortex term of $\kappa^2/\bfrho^2$,
 Eq. (\ref{8}) is directly converted  into the vortex state
\begin{equation}
\left[  -\nabla^2_\rho + \frac{\kappa^2}{\rho^2} + \rho^2  - 2 \mu +
8 \pi \xi N \phi^2(\rho) \right] \phi(\rho)= 0 \; .
\label{37}
\end{equation}
The wave function for $\kappa = 1$ vortex state is plotted in Fig. 3.
The overall shape and $N$ dependence of the vortex-state wave function
are similar with those of 3D \cite{dalf}.
As expected, the vortex state also corresponds to the $z=0$ cut of the 
anisotropic contour plot of vortex state in 3D.
We also observe that  the 2D vortex has a larger radius than the 3D one. 

The critical angular velocity or the energy difference between
the vortex state of $\kappa=1$ and the ground state $\kappa=0$ in Eq. (\ref{13})
is obtained analytically as
\begin{equation}
\vare_{\kappa=1} - \vare_{\kappa=0} = 2\sqrt{1 + \frac{\xi N}{2}} - \sqrt{1 + 2\xi N}.
\label{40}
\end{equation}
For $\vare_{\kappa=1}$, the trial function of 
$ \phi \sim \rho e^{-\rho^2}$ type was employed. 
With the increase of the number of particles, the vortex-excitation
 energy in 2D is much smaller than the 3D one, which indicates that
 vortices are expected to be produced more easily in 2D than  3D.

Although there are fundamental differences between the 2D  
and the 3D results as summarized  in TABLE 1, it will be interesting to 
compare with the quasi-2D scheme considered as a limiting case of 3D.
For comparison of the 2D trap with an extremely squeezed 3D trap,
we take the 3D external potential given by
$V_{ext}(\bfrho,z) = (1/2) m \omega^2 (\bfrho^2 + \lambda^2 z^2).$
Here $\lambda$ is the anisotropy parameter and much larger than 1 for the 
quasi-2D system, whereas $\lambda \rightarrow \infty$ for the 2D trap.  
With a fixed number of atoms, we plot the condensate wave functions
versus $\lambda$, from the 3D GPE in Fig. 4. 
We find the 3D wave functions merge very slowly
with $\lambda$, but does not reach the pure 2D limit at all.

\section{discussions}
The 2D nonlinear Schr\"{o}dinger equation(GPE) is not a simple extension 
of the 3D case but is connected into a totally different 2D collision theory
which requires different approach.
We have developed the theory of the 2D GPE for trapped 
neutral Bose atoms in a 2D harmonic trap.
Applying the quasi-2D experimental value of $\xi = 1/7$ 
to the effective interaction strength, 
one can solve the 2D GPE numerically without detailed knowledge 
of $k$ and $b$ separately.

We have obtained the stable solutions of the 2D GPE, which predicts
 possible existence of 2D BEC.
The ground-state energy of the condensate particle is also calculated 
and it is found that the 2D system becomes less stable
as the number of trapped atoms is increased,  compared with the 3D case.
We also have obtained the wave functions of the 2D vortex state numerically
and its critical angular velocity.

         The 2D BEC transition may look similar to the
Kosterlitz-Thouless(KT) \cite{kt} vortex-state transition,
but the phase transition of 2D BEC does not need any strong interaction 
between atoms.  That is the fundamental difference between
 the two transitions.  

The logarithmic dependence of the interaction potential
 on the scattering length makes the 2D system less sensitive to the species
of the atoms used for condensation.
That will make every alkali atoms of positive scattering length
show similar condensate wave patterns  in 2D.
The comparison between our 2D and well-known 3D results
 are summarized  in TABLE 1.

The quasi-2D scheme can be interpreted as a limiting case of the 3D one.
By varying the trapping field, it is possible to separate the 
single-particle states in the oscillation into well-defined bands.
To compare the situation of 2D trap with an extremely squeezed 3D trap,
we take the external potential for 3D as
$V_{ext}(r_\bot,z) = (1/2) m \omega^2 ( r_\bot^2 + \lambda^2 z^2)$,
where $\lambda \gg 1$.  
The case of negative scattering length in 2D will be discussed later.

 \acknowledgements
Authors thank G. Shlyapnikov, C. Greene, J. Macek, and H. Nha for helpful discussions.
This work was supported by the Creative Research Initiatives
of the Korean Ministry of Science and Technology.


{\footnotesize $ ^{\dagger} $ E-mail: whjhe@snu.ac.kr}
\begin{figure}
\caption{The 2D condensate wave function of neutral trapped bosonic
 atoms for  $kb = 3 \times 10^{-2}$. }
\end{figure}
\begin{figure}
\caption{The ground-state energy per particle (in unit of $\hbar \omega$) 
of Eq. (10) for  $kb = 3 \times 10^{-2}$.  }
\end{figure}
\begin{figure}
\caption{The 2D wave function of vortex state of neutral trapped bosonic
 atoms for $k b = 3 \times10^{-2}$ and a quantum circulation $\kappa =1$.}
\end{figure}
\begin{figure}
\caption{The condensate wave functions in anisotropic 3D trap for $N=100$.  
 From up to down $\lambda=2, 10, 50, 100, 500$ . 
Note that $\phi(\bfrho) = \left[ \int dz |\phi(\bfrho,z)|^2\right]^{1/2}$.}
\end{figure}
 \end{multicols}
\begin{table}
\caption{The comparison of the 3D and the 2D BEC.
The dimensionless atomic variable $\xi$ is defined by $\xi = -1/(2 \ln k b)$.}  
\end{table}
\begin{tabular}{|c|c|c|} \hline \hline
{\it Dimension} &  $D = 3$ &  $D = 2$ \\ \hline \hline
 Scattering amplitude  & $f_{3D} = - a$
 & $ F_{2D} = - i \sqrt{\frac{ 2\pi}{ k}}\xi $ \\ \hline
Interaction strength  & $g_{3D} = \frac{4\pi \hbar^2 a}{m}$
& $g_{2D} =\frac{4 \pi \hbar^2 \xi}{m} $ \\ \hline
Healing length & $\xi_{3D} = (8 \pi n_{3D} a)^{-1/2}$
& $\xi_{2D} = \left(8 \pi \xi n_{2D}\right)^{-1/2}$
 \\ \hline
Chemical potential &\ &\ \\ (noninteracting)
& $\mu_{3D}=1.5\,\hbar \omega $ & $\mu_{2D}= \hbar \omega$ \\
(strongly interacting)
&$\mu_{3D}=\frac{1}{2}\left(\frac{15 a N}{a_{h.o}}\right)^{2/5}\hbar \omega$
&$\mu_{2D}= \left( 4 \xi N\right)^{1/2}\hbar\omega$
 \\ \hline
Radius of condensation & $r_c = \left(\frac{15 a N}{a_{ho}} \right)^{1/5}$
& $\rho_c = \left(16 \xi N \right)^{1/4}$ \\ \hline
Ground state energy & $E_{3D} = \frac{5}{7}\mu_{3D} N \propto N^{7/5}$
        & $E_{2D}=\frac{2}{3}\mu_{2D} N \propto N^{3/2} $ \\ \hline
Center lowering of &\ &\ \\ density profile
 & $|\phi_{3D}(0)|^2 \propto N^{-3/5}$
        & $|\phi_{2D}(0)|^2 \propto N^{-1/2}$ \\ \hline
 \hline
\end{tabular}
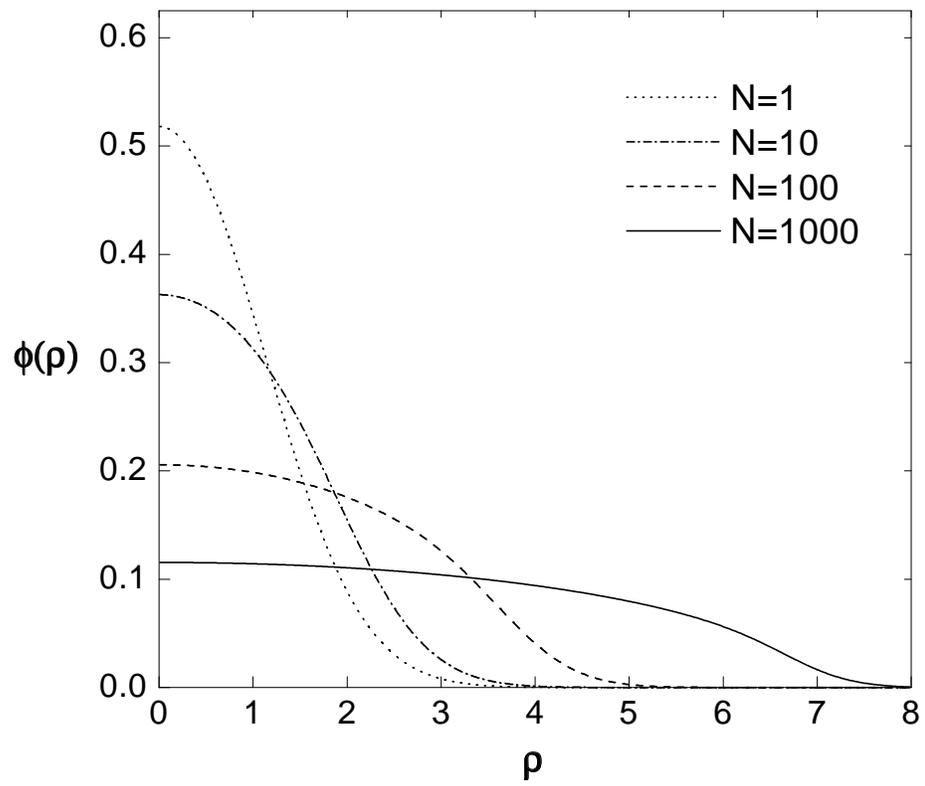
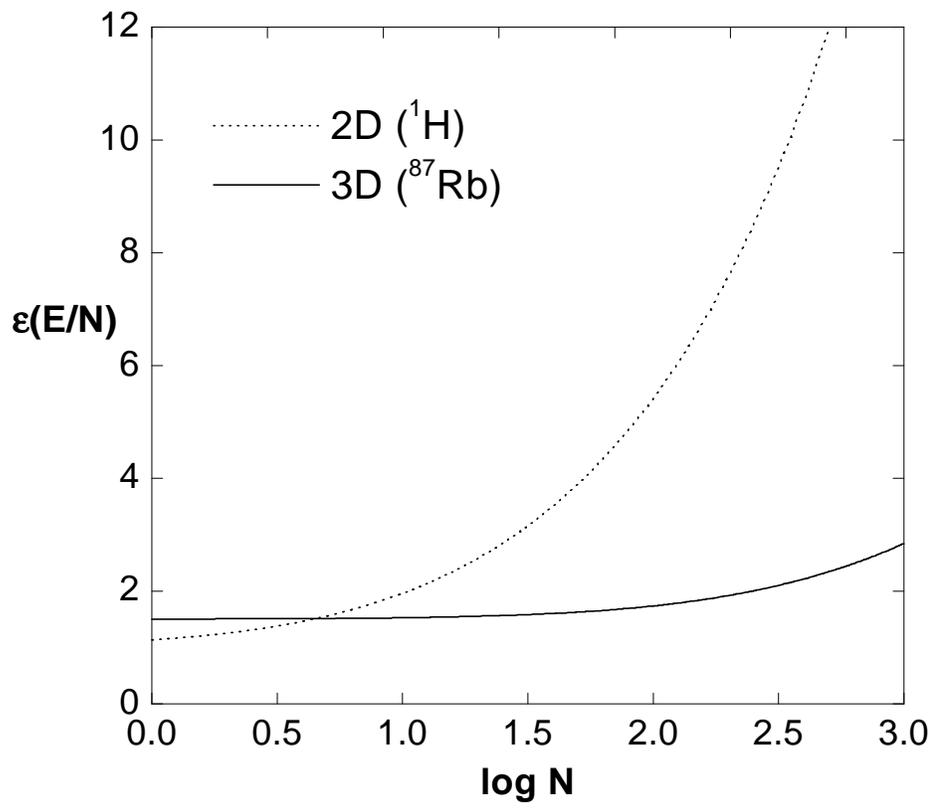
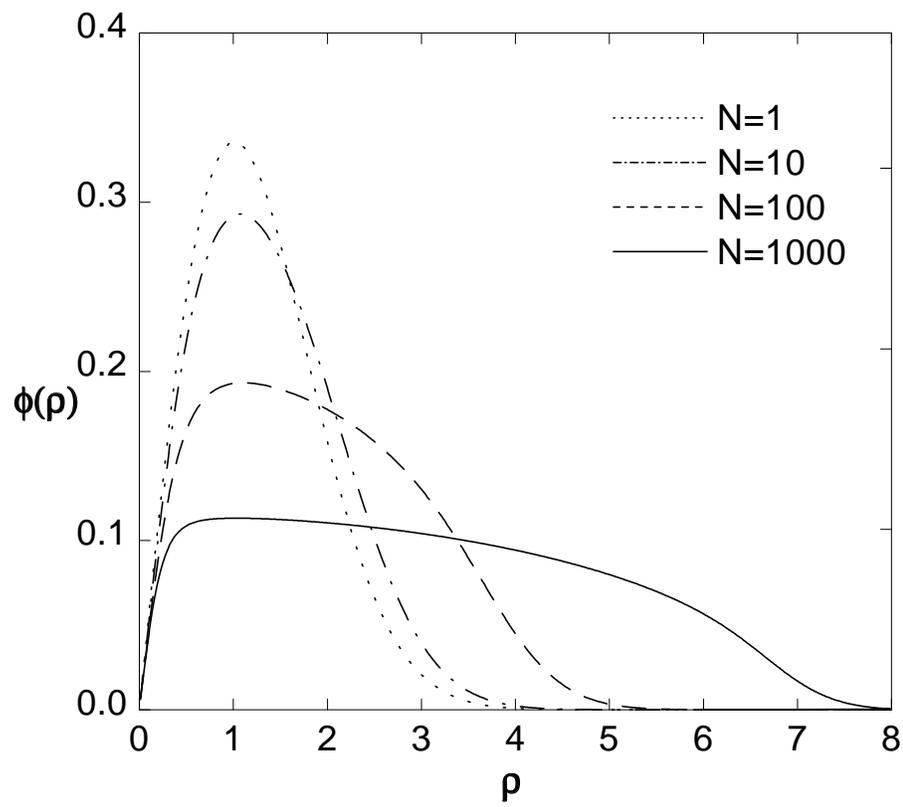
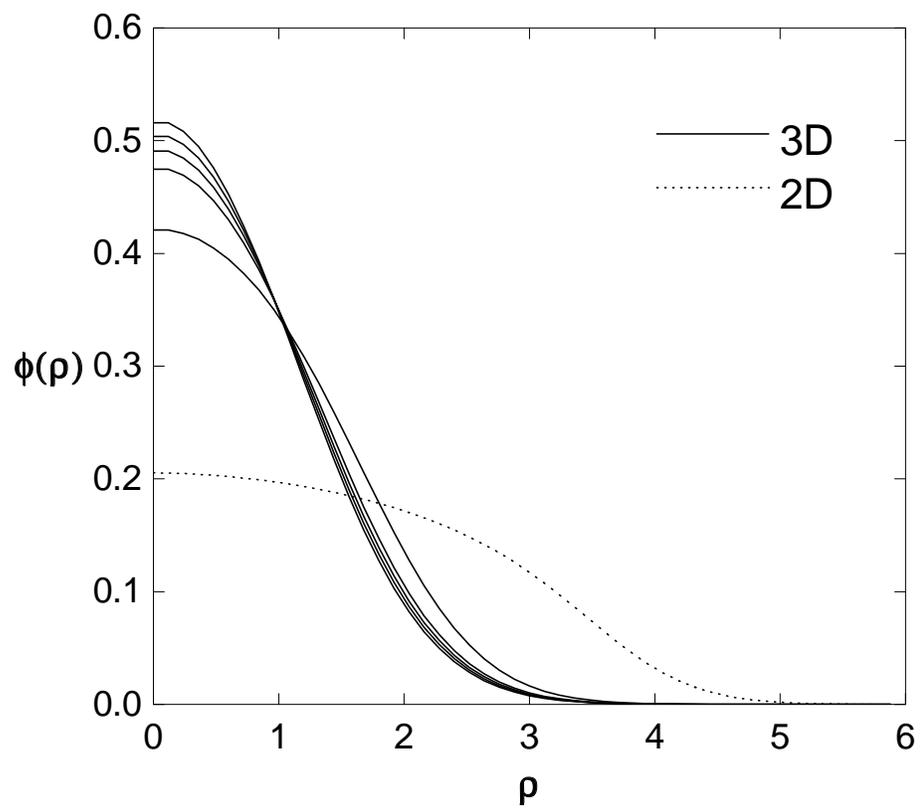
\end{document}